\documentclass[12pt]{article}
\usepackage{graphicx}
\usepackage{amsmath}
\usepackage{amssymb}
\newcommand{\be}{\begin{equation}}
\newcommand{\ee}{\end{equation}}
\newcommand{\bea}{\begin{eqnarray}}
\newcommand{\eea}{\end{eqnarray}}
\newcommand{\pr}{\partial}

\newcommand{\bse}{\begin{subequations}}
\newcommand{\ese}{\end{subequations}}

\begin{document}
\title{Domain walls in Born-Infeld-dilaton background}
\author{Debaprasad Maity\footnote{E-mail: debu@imsc.res.in}\\
The Institute of Mathematical Sciences,\\ 
C.I.T. Campus, Taramani,\\
Chennai - 600 113, India}
\maketitle
\begin{abstract}
We study the dynamics of domain walls
in Einstein-Born-Infeld-dilaton theory. Dilaton is  
non-trivially coupled with the Born-Infeld electromagnetic field. 
We find three different types of solutions consistent with
the dynamic domain walls. 
For every case, the solutions have singularity. 
Further more, in these backgrounds, we study the 
dynamics of domain walls. We qualitatively 
plot various  form of the bulk metrics
and the potential encountered by the domain walls. 
In many cases, depending upon the value of the parameters, 
the domain walls show bouncing universe and also undergo
inflationary phase followed by standard decelerated expansion. 

\end{abstract}
\section{Introduction}\label{intro}
Our universe as a four dimensional domain wall [1-9]
in an extra dimensional spacetime is an interesting field to study
particularly in
the cosmological context. The domain wall
is allowed to move in an extra spacelike dimension. The inherent nexus
between this dynamics of the domain wall 
and Hubble like expansion equation of its scale factor attracts
much attention to study
the cosmology in a new perspective. In this scenario, 
all the standard model fields are assumed to 
be either in the bulk but dynamically peaked at the position
of the wall \cite{rubakov,dvali} or \cite{polchinski} fully 
localized on the domain wall depending upon what kind of 
model we are going to construct.

As we know for the past few years, domain wall are actively
considered to embed as a four dimensional Friedmann-Robertson-Walker
(FRW) universe in bulk spacetime with various possible matter
components\cite{kraus,csaki}. 
Important aspect of this embedding, as just stated, is that
boundary condition so called Israel junction condition\cite{israel} 
across the wall actually leads to Hubble expansion equation.
So, what it turns out that various parameters of the 
bulk solutions effectively act as 
(invisible) energy density with different equations of 
states on the wall. So, by tuning these various bulk
parameters in a model under consideration, one can in principle construct
viable cosmology. In fact, on the domain wall there exits  
bouncing cosmology. The important feature to mention in this kind of
cosmology is non-singular transition between a
contracting phase of the scale factor of the wall, 
and a following expanding stage
\cite{sudipto,novello}.
In this report will not going to construct the cosmological model.
We will first try to find out the various possible background 
configurations in a string inspired theoretical model
and then do a qualitative analysis on the dynamic domain walls in those
backgrounds. 

As is well known, Born-Infield type generalisation of 
abelian and non-abelian gauge field theories have
long been the subject of interest in the context of 
superstring theory. For the abelian case, it was first
argued in the reference \cite{teys} to be an open string effective action
taking into account all order loop corrections in $\alpha'$(string coupling).
On the other hand, it has also been noted that the D-branes \cite{leigh} and 
some soliton solutions of supergravity, are governed by the 
Born-Infeld action. For various reasons particularly in 
the context of ADS/CFT correspondence, extending the gravitational
background by including Born-Infeld gauge field in addition to 
Einstein-Hilbert term, has recently been
considered extensively in \cite{tdey,Born-infeld}.
In a slightly different direction, we will try to analyse 
the dynamics of the domain wall in these background.

So, as a follow up of \cite{debu}, here we will be  
discussing on dynamic domain walls in the more general 
Einstein-Born-Infield-dilaton \cite{dil} 
background along the line of \cite{chamblin}.
String theories in their low-energy supergravity 
limit gives rise to effective models of gravity in higher dimensions which 
involve an infinite series of higher curvature terms in the gravity as
well as gauge sector. 
For our purpose in this report, we will ignore 
the higher spacetime curvature terms. Although including the same
is very much worth calculating. We leave this for future study 
(recent attempt towards this direction \cite{0806.2481}).
Considering this born-Infield Lagrangian, 
there has been considerable work on understanding the role 
of these higher derivative gauge field 
in various points of view, especially with
regard to black hole physics \cite{tdey,Born-infeld}. 
In this report we will analyze various static black hole
as well as non-static spacetime solutions in consistent with
dynamic domain walls in the aforementioned string inspired model.
However, in addition to dilaton scalar field coupled with the 
BI Lagrangian in our theory, we also assume the dilaton to 
be coupled with domain
wall in an exponential way \cite{chamblin}.
We first analytically find
three different types of metric solutions under specific
relations among the various constant parameters in the theory.
We have taken into account full back-reaction of the
domain wall on background spacetime. 
Subsequently we study the structure of the
solutions in details by plotting them in various
possible region of the parameter space. 
In many cases, we find static black hole solution as well as 
various non-static solutions depending upon the choice of
parameters.
Then, we study the dynamics of a domain wall in those
bulk backgrounds.
A topic of particular interest in these kind
of scenarios is how inflation occurs on the wall.
In this regard, we find, for a
wide range of parameter space of the model under
consideration, the domain wall indeed inflates either in the
early stage of the evolution followed by
standard decelerated expansion or in asymptotic limit of its scale factor. 
Depending upon the choice of parameters, 
the inflation is either exponential or power law type with respect to
the domain wall proper time.
One important feature in these kind of scenarios is natural
emergence of inflation as
well as decelerated expansion phase on the domain wall world volume.
Various energy densities which drive this dynamics 
, strictly come from the bulk.

We structured this report as follows: In the section \ref{sec1}, we
start with an action corresponding to a domain wall
moving in Born-Infield-dilaton background. We explicitly 
write down the equations of motion and corresponding boundary conditions
across the domain wall. In section \ref{sec2}, by taking static
metric ansatz, we analytically solve the Einstein equations. 
We get three different types of solutions depending upon the 
value of the parameters in our theory.
Due to complicated expressions, we study the structure of these solutions 
by graphical representation in various limits of radial coordinate. 
In section \ref{sec3}, again following the line of \cite{chamblin,debu}, 
we plot the various potentials encountered by the domain walls and 
qualitatively study their dynamics. 
There are several situations we find, for which domain
wall undergoes an inflationary phase as well as standard decelerated 
expansion. We also get bounce for finite value of scale factor 
as well as periodic universe on the domain wall world volume. 
This might lead to a hope of constructing the cosmological 
model in this theory also.
Finally, in the section \ref{con}, we do some concluding remarks and 
describe some futures directions to work.

\section{Einstein equations and Boundary conditions} \label{sec1} 
We start with an string inspired action of Einstein-Born-Infeld
system in an arbitrary spacetime dimension $n$. 
Along with this there exists a bulk scalar field $\phi$ 
called dilaton. The dilaton is assumed to 
have nontrivial coupling with the Born-Infeld field 
$A_{B}$ and a co-dimension one domain wall.  
The action takes the from 
\begin{equation} \label{action}
S = \int d^n x\sqrt{-g}\left( \frac 1 2 R ~-~ \frac 1 2 
\pr_{A} \phi \pr^{A} \phi ~-~ V(\phi) ~+~ {\cal L}(F,\phi)\right) ~+~ S_{DW},
\end{equation}
where action for the domain wall is
\begin{center}
$S_{DW} ~=~  ~-~ \int d^{n-1} x \sqrt{-h} \left(
\{K\} + \bar{V}(\phi) \right)$.
\end{center}
The expression for ${\cal L}(F,\phi)$  is 
\begin{equation}
{ L(F,\phi)} = {4{\lambda}^2 e^{2 \gamma \phi} \Big (1-\sqrt{1+\frac{e^{-4 \gamma \phi} F^{AB} 
F_{AB}}{2{\lambda}^2}}\Big)}.
\label{}
\end{equation}
The constant $\lambda$ is the Born-Infeld parameter and has
the dimension of mass. In the limit $\lambda \rightarrow \infty $,  $ L(F)$ 
reduces to the standard Maxwell form with a scalar field coupled exponentially like
\begin{equation}
{ L(F)} = {- e^{-2 \gamma \phi} F^{AB} F_{AB}} +{\cal{O}}(F^4).
\label{}
\end{equation}
$h_{AB}$ is the induced metric on the domain wall. $K$ is the trace of the
extrinsic curvature of the domain wall.
For simplicity, in this paper, we will work with the convention 
that $16 \pi G =1$, where $G$ is the Newton's 
constant.

By varying the action with respect to the gauge field $ A_B$, dilaton field
$\phi$ and the metric $ g_{AB} $, we get the equations
of motion as
\bea
&&R_{AB} = \pr_{A} \phi \pr^A \phi + \frac 2 {n -2} V(\phi) g_{AB}+  \frac {8 \lambda^2} {(n-2)}
e^{2 \gamma \phi}\left\{2 {\cal Y} \frac {\pr {\cal L}}{\pr {\cal Y}} - {\cal Y} \right\} g_{AB}
- 8 e^{- 2 \gamma \phi} \frac {\pr {\cal L}}{\pr {\cal Y}} F_{AC}F_B^C ~~~~\\
&&D_C \pr^C \phi - \frac {\pr (\phi)}{\pr \phi} + 8 \lambda^2 \gamma  e^{2 \gamma \phi}\left\{2 {\cal Y} \frac {\pr {\cal L}}{\pr {\cal Y}} - {\cal Y} \right\} = 0 \\
&& D_A\left( e^{- 2 \gamma \phi} \frac {\pr {\cal L}}{\pr {\cal Y}} F^{AB}\right) = 0 
\eea
where, $D_A$ is co-variant derivative with respect to the bulk 
metric and the new variable ${\cal Y} = \frac{e^{-4 \gamma \phi} F^{AB} 
F_{AB}}{2{\lambda}^2}$. We also have the corresponding boundary 
conditions as follows
\bea \label{bond}
&&\{K_{MN}\} = - \frac 1 {n -2} \bar V(\phi) h_{MN} \\
&&\{n^{M}\pr_{M} \phi\} = \frac {\pr \bar{V}(\phi)}{\pr \phi}
\eea
where, $n^M$ is the unit normal to the domain wall.
$R$ is the curvature scalar. The first boundary condition comes
from the Israel junction condition (for details \cite{chamblin})
across the wall.

We will be considering the solution which has
reflection symmetry($Z_2$) across the domain wall. 
So, from Eq.\ref{bond}, expression
for the extrinsic curvature turns out to be
\be
K_{MN} = - \frac 1 {2(n -2)} \bar V(\phi) h_{MN} 
\ee
We consider the static spherically symmetric bulk metric ansatz as
\be \label{metric}
ds^2 = - N(r) dt^2 + \frac 1 {N(r)} dr^2 + R(r)^2 d\Omega_{\kappa}^2
\ee
where, we have taken $ d\Omega_{\kappa}^2$ as the line element 
on a $(n -2)$ dimensional space of constant curvature with the metric
$\tilde {g}_{ij}$. The Ricci curvature of this sub-space is 
$\tilde {R}_{ij} = k (n - 3) \tilde {g}_{ij}$ with $ k \in \{-1,0,1\}$

We want to get the spherically symmetric solutions corresponding to a 
homogeneous and isotropic induced metric on the domain wall like
\bea
ds_{wall}^2 = - d\tau^2 + R(\tau)^2 d\Omega_{\kappa}^2,
\eea
which is Robertson-Walker metric. $\tau$ is the
domain wall proper time. 
The size of the $n$-dimensional domain wall universe is 
determined by the radial distance,
$R$, which in turn determines the position of it in the bulk spacetime.

However, by using the unit normal pointing into $r < r(t)$ and 
the unit tangent to the moving wall 
\bea
&&n_M = \frac {\sqrt{N}} {\sqrt{N^2 - \dot{r}^2}}(\dot{r}, -1, 0,\dots,0), \\
&&u^{M} = \frac {\sqrt{N}} {\sqrt{N^2 - \dot{r}^2}}(1, \dot{r}, 0,\dots,0),
\eea
respectively,
one can express the induced
metric on the domain wall and its extrinsic curvature as
\bea
&&h_{MN} = g_{MN} - n_{M} n_{N} \\
&&K_{MN} = h_M^P h_N^Q \nabla_P n_Q.
\eea
Where $\dot{r} = \frac {\pr r}{\pr t}$.

So, the expressions for the components of the extrinsic curvature 
come out to be
\begin{subequations} \label{excomp1}
\bea 
&&K_{ij} = -  \frac {R'}{R} \frac {N^{3/2}}{\sqrt{N^2 - \dot{r}^2}} h_{ij} \\
&&K_{00} = \frac 1 {\dot r} \frac d {dt} 
\left( \frac {N^{3/2}}{\sqrt{N^2 - \dot{r}^2}}\right) .
\eea
\end{subequations}
'Prime' is derivative with respect to bulk radial coordinate $r$.

Using the equations for $K_{ij}$ into $K_{00}$, and then integrating
one gets
\bea \label{E1}
R' = C \bar{V}(\phi) .
\eea
Now, by using Eq.\ref{E1} in the 
boundary condition for the scalar field one gets
\bea \label{E2}
\frac {\pr \phi}{\pr R} = - \frac {n -2} R \frac 1 {\bar V} \frac {\pr \bar V}
{\pr \phi}
\eea
This equation has to hold at every point in the bulk visited by the 
domain wall. So, if the wall visits a range of R, then the above equation
can be solved to yield $\phi$ as a function of R without specifying
the bulk potential. This gives us a consistency condition for the 
dynamic domain wall coupled with the bulk scalar to exists. 
In what follows, we will use this condition to get final
solution for the scalar field and the metric.

\section{The solutions for bulk metric} \label{sec2}
In this section, we first solve the equation for Born-Infeld field. 
Then using this solution to the remaining equations of 
motion we will find out the solution for metric under 
static bulk metric ansatz Eq.\ref{metric}. 

So, we note that a class of solutions for the equation of 
Born-Infield electromagnetic field 
can be written down with all the components of $F^{AB}$ being
zero except $F^{rt}$. The solution looks like
\bea \label{BIsol}
F^{rt} = \frac {2 Q \lambda e^{2 \gamma \phi}}{\sqrt{4 Q^2 + \lambda^2 R^{2 n -4}}}
\eea
where, $Q$ is the integration constant and 
related to the electromagnetic charge. One can define 
electromagnetic charge with respect to an asymptotic observer as
\bea
q = \frac 1 {4\pi} \int_{\Sigma_\infty} e^{- 2 \gamma \phi}~ {^*F} = 
\frac {Q \omega_{n-1}}{4 \pi},
\eea
where, ${^*F}_{AB} = \frac 1 {2 \sqrt{-g}} 
\varepsilon^{ABCD} F^{CD}$ and $\Sigma_{\infty}$ is a hyper-surface
at $R \rightarrow \infty$. $\omega_{n-1}$ is volume of unity $n$ sphere.
One can notice that the electric field is finite at $R = 0$. 
This is expected in Born-Infeld theories.
  
Now using Eq.\ref{BIsol} and the metric ansatz Eq.\ref{metric}, 
one can read out the remaining equations of motion  as
\begin{subequations}
\bea
&&\frac {R''} {R} = - \frac 1 {n-2} \phi'^2\\
&& \frac 1 {2 R^{n-2}} \left\{N \left(  R^{n-2}\right)'\right\}' -
\frac {k(n-3)(n-2)} {2 R^2} = - V - {\cal T}_{22}(R,Q) \\
&& \frac {n-2}{4 R^{n-2}} \left( N' R^{n-2}\right)' = - V - {\cal T}_{00}(R,Q) \\
&&\frac 1 {R^{n-2}}\left( \phi' N R^{n-2}\right)' = 
\frac {\pr V(\phi)}{\pr \phi} + 8 \lambda^2 \gamma e^{2 \gamma \phi}  {\cal F}(r,Q),
\eea\end{subequations}

where, various new notations are given below,
\bea
{\cal T}_{22}(R,Q) = 4 \lambda^2  e^{2 \gamma \phi} {\cal F}(R,Q)~~;~~{\cal T}_{00}(R,Q) =
4(n-2) \lambda^2  e^{2 \gamma \phi}\left[\frac {{\cal F}(R,Q)}{n-2} + \frac {{\cal G}(R,Q)}2 \right]
\eea
and
\bea
{\cal F}(R,Q) = \frac {\sqrt{4 Q^2 + \lambda^2 R^{2 n -4}}}{\lambda R^{n -2}} - 1~~~;~~~
{\cal G}(R,Q) = - \frac {4 Q^2}{\sqrt{4 Q^2 + \lambda^2 R^{2 n -4}}} \frac 1 {\lambda R^{n -2}}.
\eea

${\cal T}_{00}$ and ${\cal T}_{22}$ are proportional to the $tt$  and spatial
components of the energy-momentum tensor for the Born-Infield Lagrangian respectively.

Now, in order to find the solutions of Einstein equation, we will employ
the Eqs.(\ref{E1},\ref{E2}). So, taking the Liouville
type brane potential 
\be
\bar V(\phi) = {\bar V}_0 e^{\alpha \phi},
\ee
one can easily get the solution for the scalar field
without specifying the bulk potential, as well as 
for the radius $R(r)$ of the unit sphere $\Omega_k$ as
\begin{subequations} \label{sol}
\bea 
&&\phi = \phi_0 - \frac {\alpha (n-2)}{\alpha^2(n-2) + 1} log (r)\\
&&R(r) = C {\bar V}_0 e^{\alpha \phi_0} 
r^{\frac 1 {\alpha^2(n-2) + 1}},
\eea
\end{subequations}
where $\phi_0$ and $C$ are integration constants.
Furthermore, in order to have the solution for the bulk metric, we need to specify the 
bulk potential. So, again we take the same Liouville type bulk potential,
\be
V(\phi) = V_0 e^{\beta \phi}
\ee
where, $V_0$ is constant.
Now, by considering above two expressions Eqs.(\ref{sol}) 
for $R$ and $\phi$ as solutions ansatz
and making use of the bulk potential for the scalar field, we find three 
different types of solutions corresponding to the full Einstein
equations of motion. 
In what follows, we discuss about
the nature of these various types of solutions and subsequently 
the dynamics of the
domain walls in those bulk backgrounds.

{\bf Type-I solution}: When, $\alpha = \beta = \gamma = 0$. We note that the 
bulk and brane potential play the roll of cosmological constant 
and brane tension respectively. So, effectively, the action is a
Einstein-Born-Infeld system with a bulk cosmological constant and domain
wall with fixed tension.

By choosing these particular set of value of the parameters, 
the solution turns out
to be
\bea
N(r) &=& k - 2 M r^{-(n-3)} - \left(\frac {2 V_0}{(n-2)(n-1)} - \frac {8 \lambda^2}{(n-2)(n-1)}\right)r^2 \\ 
&+& \frac {8 \lambda r^{-(n-4)}} {(n-1)(n-2)} \left( {- \sqrt{4 Q^2 + \lambda^2 r^{2n -4}}}   
+  \frac {4 (n-2) Q^2 r^{-(n-2)}}{\lambda (n-3)}~~ {\cal D}(r,Q) \right)\\
&&R(r) = r ~~~~;~~~ \phi = \phi_0 , 
\eea
Where  $M$ and $\phi_0$ are integration constants and
\bea 
{\cal D}(r,Q) = {_2}F_1\left[\frac {n-3} {2n -4}, \frac 1 2 , \frac {3n-7}{2n -4}, -\frac {4 Q^2 r^{-(2n -4)}}
{\lambda^2}\right].
\eea
Now, as a whole expression, the solution looks much difficult. 
So, it is rather easer to see the metric in various limits of 
the radial coordinate and study its behaviour.

   As we note, for large r the expression for the above solution looks like
\bea
{N(r)|_{r\rightarrow \infty}} = k - \frac {2 V_0}{(n-2)(n-1)} r^2 - 2 M r^{-(n - 3)} +
\frac {16 Q^2}{(n-3)(n-2)}r^{-(2n-6)} + {\cal O}(r^{\frac{10 -4n}{1+c^2}}),
\eea
and for small $r$ limit, 
\bea
 {N(r)|_{r\rightarrow 0}} = k - 2 {\cal M}_1 r^{-(n-3)} 
&-& \frac {16 \lambda {\cal Q}_1} {(n-1)(n - 2)} r^{-(n-4)} - 
\frac {2 {\cal V}_1 }{(n-1)(n-2)} 
r^{2} \nonumber \\ &-& \frac {8 \lambda {\cal H}_1} {(n-1)(n - 2)}
 r^n + {\cal O}(r^{{3n-4}}),
\eea
where
\bea 
{\cal M}_1 & =& M - \frac {16 Q^2 \Gamma[\frac {3n-7}{2n -4}]\Gamma[\frac {1}{2n-4}]}{\sqrt{\pi}(n-1)(n-3)}
 \left(\frac {4 Q^2}{\lambda^2} \right)^{-\frac {n-3}{2n -4}}\\
{\cal Q}_1 &=& Q - \frac {2 Q^2 (n-2) \Gamma[\frac {3n-7}{2n -4}]\Gamma[\frac {-1}{2n-4}]}{\lambda (n-3) 
\Gamma[\frac {n-3}{2n -4}] \Gamma[\frac {2n -5}{2n -4}]} \left(\frac {4 Q^2}{\lambda^2} 
\right)^{-\frac 1 2} \\
{\cal H}_1 &=& \frac {\lambda^2}{4 Q} + \frac {1}{(2n-3)}\left(\frac {4 Q^2}{\lambda^2} 
\right)^{-1} (Q - {\cal Q}_1) \\
{\cal V}_1 &=& = V_0 - 4 \lambda^2 .
\eea

The solution for the scalar field becomes constant \cite{tdey}. 
In order to have better understanding of the metric, 
we have plotted the metric in the Fig.1 for several possibilities of
parameter values. Now it is easy to read off the horizon structure 
from the figure. As is seen from the figures that for every case, 
there exists singularity at $r = 0$ which is either timelike or spacelike.
   
 We have four possibilities for different region of the parameter 
space $(V_0 , M )$. When $V_0 > 0, M > 0$, for the first case the 
metric would be either a Reisner-Nordstrom(RN)
black hole inside the cosmological horizon ($k = 1$) or 
only de Sitter space ($k = 0,-1$). 
For the other case, we have either Schwarzschild-de Sitter black hole 
which has black hole ($k=1$) and cosmological horizon or non-static
spacetime ($k = 0,-1$) with a naked singularity. As is clear, for every case, 
the behaviour of the metric near the singularity region $r\rightarrow 0$ is 
governed by the sign of ${\cal M}_1$ . In all these cases asymptotically, 
the metric becomes de-sitter space where, $V_0$ is playing the roll of 
cosmological constant. Usually, the ADM \cite{abbot} mass of 
the black hole is taken to be
\bea
m_{ADM} = \frac {2 n\omega_{n-1}} 2 M,
\eea 
where, $\omega_{n-1}$ is the volume of the unit $n$-sphere.

When $V_0 > 0, M < 0$, the singularity is either space-like or 
time like defined by the sign of ${\cal M}_1$ , negative or 
positive respectively. The interplay between the value of $ Q$ and
$M$ determines the singularity structure at $r = 0$. 
Asymptotically the metric is Anti-de Sitter. 
For one case we have Reisner-Nordstrom (RN) black hole or 
there is a naked singularity at $r=0$. For the other case, 
the spacetime is Schwarzschild-anti-de Sitter.

If $V_0 < 0, M > 0$, for every case $k = 0,\pm 1$, 
the metric has cosmological horizon with
asymptotically de Sitter space.

When $V_0 < 0, M < 0$, for $k = 0,1$, the metric has a 
time like naked singularity at $r = 0$ for any value of the 
other parameters present. However, $k = 1$ leads to a possibility
of having a RN black hole in an asymptotically Anti-de Sitter 
spacetime for certain range of parameter space $(V_0 , M)$ and $\lambda Q^2$.
\begin{figure}
\includegraphics[width=5.5in,height=1.5in]{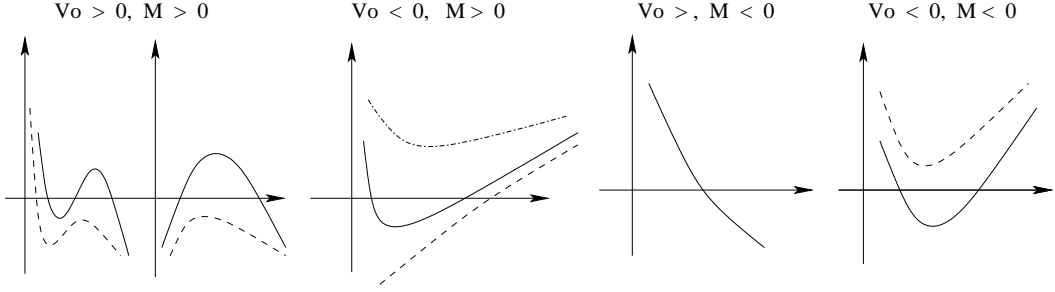}
\caption{N(r) for type-I solution.} \label{one}
\end{figure}

\begin{figure}
\includegraphics[width=5.50in,height=3.2in]{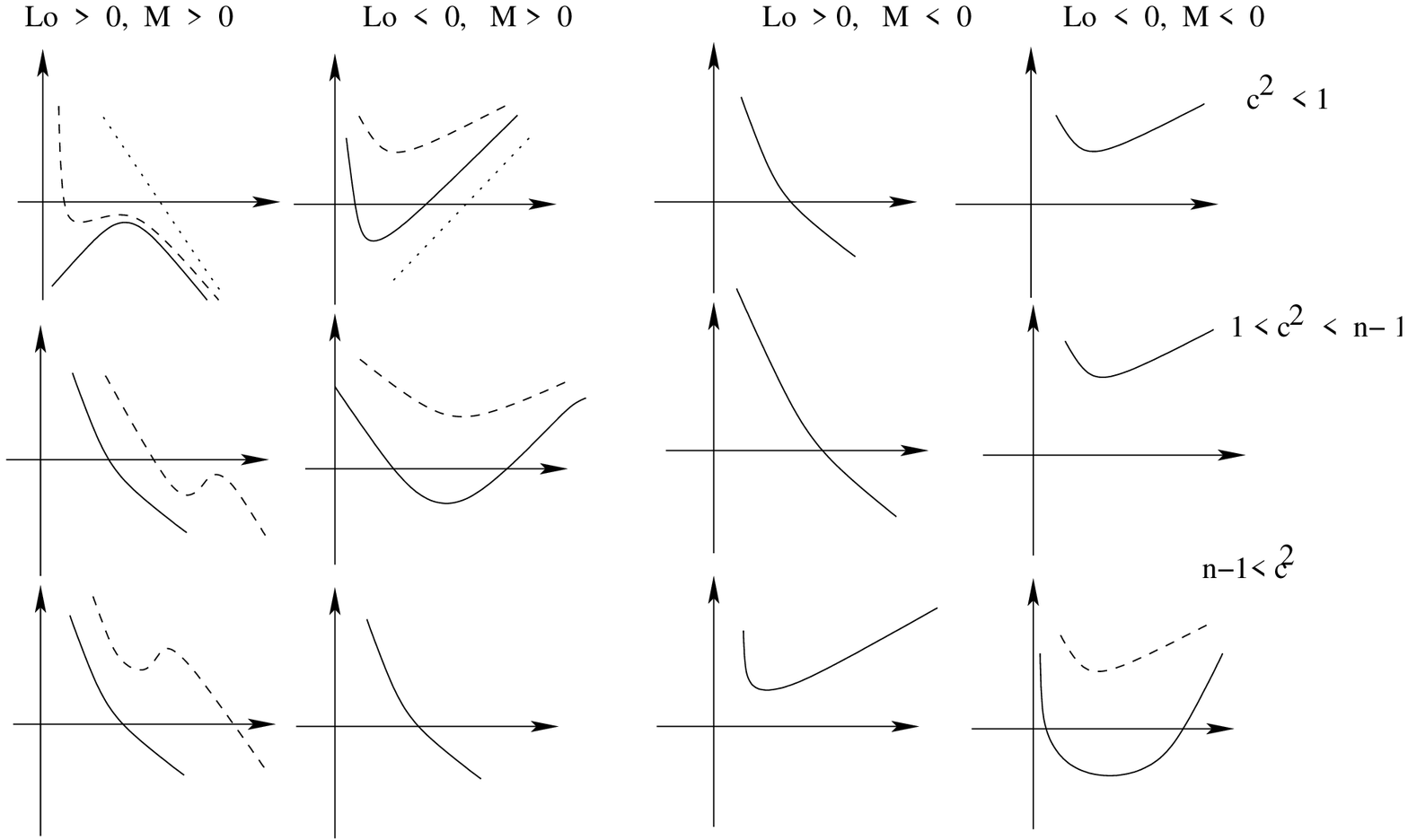}
\caption{N(r) for type-II solution. 
} \label{two}
\end{figure}

{\bf Type-II solution}: For $\alpha = \frac {\beta} 2 =  \gamma ~;~ 
k = 0$, which means that metric with flat spatial section is the only
allowed configuration for this choice of parameters. The 
solution comes out to be
\bea
N(r) &=& -(1 + c^2)^2 r^{\frac 2 {1 + c^2}} \left[\frac 1 {(n-1-c^2)} \left\{ {2 L_0} - 
\frac {8 \lambda^2 \Omega}{(n-2)(1+c^2)^2}\right\} + 2 M r^{- \frac {n - 1 - c^2}{1 + c^2}} \right] \\ 
&& + \frac {8 \lambda \Omega r^{-\frac {n-4}{1 + c^2}}}  
{(n-2)(1 + c^2 -n)} \left(\sqrt{4 Q^2 + \lambda^2 r^{\frac {2n -4}{1 +c^2}}} -
\frac {4(n -2)Q^2 r^{-\frac {n-2}{1 + c^2}}}{\lambda (- 3 + c^2 +n)}~ {\cal P}(r,Q)\right)\\
&&R(r) = r^{\frac 1 {1 + c^2}} ~~~;~~~ \phi(r) = \sqrt{n-2}\left(
\phi^*_0 - \frac b {1 + c^2} log (r) \right),
\eea
where $\phi^*_0 = \phi_0/\sqrt{n-2}$, the integration constants. 
Various other notations are given below,
\be
c = \frac 1 2 \beta \sqrt{n-2}~~~~;~~~~ L_0 = \frac 
{V_0 e^{2 c \phi^*_0}} {n-2}~~~~;~~~~\Omega = 
\frac {(1 + c^2)^2 e^{2 c \phi^*_0}}{n-2}
\ee
and
\bea
{\cal P}(r,Q) = {_2F_1}\left[\frac {c^2 + n -3}{2n-4},\frac 1 2,\frac{3n- 7 + c^2}{2n-4},-\frac{4 Q^2 
r^{-\frac {2n-4}{1 + c^2}}} {\lambda^2} \right].
\eea

Again, taking the large $r$ limit, the above solution turns out to be
\bea
\frac {N(r)|_{r\rightarrow \infty}}{(1 + c^2)^2} = r^{\frac 2 {1 + c^2}} \left[\frac {2 L_0} {(1+c^2 -n)}  - 2 M r^{- \frac {n - 1 - c^2}{1 + c^2}} \right] +
\frac {16 \Omega' Q^2}{(n-3+c^2)}r^{-\frac {2n-6}{1 +c^2}}+ {\cal O}(r^{\frac{10 -4n}{1+c^2}}),
\eea
and for small $r$ limit, 
\bea
\frac {N(r)|_{r\rightarrow 0}}{(1+c^2)^2} = - 2 {\cal M}_2 
r^{-\frac {n-3-c^2}{1 +c^2}} 
&-& \frac {16 \lambda \Omega' {\cal Q}_2} {(n-1 - c^2 )} 
r^{-\frac{n-4}{1 +c^2}} - \frac {2 {\cal V}_2}{n-1-c^2} 
r^{\frac 2{1 +c^2}} \nonumber \\ &-& \frac {8 \lambda \Omega' {\cal H}_2} 
{(n-1 - c^2 )}
 r^{\frac n {1 +c^2}}  + {\cal O}(r^{\frac{3n-4}{1+c^2}}),
\eea
where
\bea 
{\cal M}_2 & =& M - \frac {16 Q^2 \Omega' (n-2)\Gamma[\frac {3n-7+c^2}{2n -4}]\Gamma[\frac {1-c^2}{2n-4}]}{\sqrt{\pi}(n-1-c^2)(n-3+c^2)} \left(\frac {4 Q^2}{\lambda^2} \right)^{-\frac {n-3+c^2}{2n -4}}\\
{\cal Q}_2 &=& Q - \frac { Q (n-2) \Gamma[\frac {3n-7+c^2}{2n -4}]\Gamma[\frac {c^2-1}{2n-4}]}{(n-3+c^2) 
\Gamma[\frac {n-3+c^2}{2n -4}] \Gamma[\frac {2n -5 + c^2}{2n -4}]} \\
{\cal H}_2 &=& \frac {\lambda^2}{4 Q} + \frac {(1-c^2)}{(2n-3-c^2)}\left(\frac {4 Q^2}{\lambda^2} 
\right)^{-1} (Q - {\cal Q}_2) \\
{\cal V}_2 &=& = L_0 - 4 \lambda^2 \Omega'\\
\Omega' &=& \frac {e^{2 c \phi^*_0}}{n-2}.
\eea

For this case again, we have plotted  
various possibilities for different values of the parameters present in the 
expression for $N(r)$.
All the detailed structure can easily be
read off from the corresponding Fig.\ref{two}.

The structure of the space time is depending upon the value of $c$. 
If $c^2 < 1$, the singularity behaviour is determined by the sign of 
${\cal M}_2$. As is seen from the expression of ${\cal M}_1$, for $M < 0$,
it is always negative which leads to timelike singularity at $r = 0$.
Whereas for $M >0$, the metric can have both types of singularities
depending upon the value of other parameters $Q$ and $\lambda$. 
This leads us to a quite distinct singularity behaviour which is expected
in a Born-Infeld theory as opposed
to the standard RN black hole spacetime in Einstein-Maxwell theory.
In particular, the spacetime singularity comes from a mass dependent
term in the metric.

On the other hand, for $c^2 > 1$, ${\cal Q}_2$ determines the singularity
behaviour. It is interesting to note that the expression for ${\cal Q}_2$
is such, the combined expression $\frac {{\cal Q}_2} {(n-1 - c^2 )}$
is always negative. This in turn leads to timelike
singularity for any value of $c^2 >1$ as is also clear from the
plots Fig.\ref{two}.

However, for a wide range of parameter space, the metric has a 
cosmological horizon. For $c^2 < 1$, and $1 < c^2 < n - 1$, 
asymptotic structure of the spacetime is depending upon the value of
$L_0$ which is playing the roll of cosmological constant. 
On the other hand for $c^2 > n - 1$, it is governed by the sign of 
the parameter $M$ . As we note the asymptotic expression for the metric 
to be
\bea \label{first}
ds^2 &=& r^{\frac 2 {1 +c^2}}\left( - L_0 dt^2 + d\Omega_k^2\right)
+ \frac 1 {L_0 r^{\frac 2 {1 +c^2}}} dr^2 ~~~~~~~~ \hbox{for $c^2 < n-1$}\\
\label{second}
ds^2 &=& r^{\frac 2 {1 +c^2}}\left( - h(r) dt^2 + d\Omega_k^2\right)
+ \frac 1 {h(r) r^{\frac 2 {1 +c^2}}} dr^2 ~~~~ \hbox{for $c^2 > n-1$},
\eea
where
\bea
h(r) = \left( \frac {2 L_0}{n-1-c^2} + 2 M r^{-\frac {n-1-c^2} {1 +c^2}}\right).
\eea
So, for the second case, we observe that the spacetime is 
asymptotically ADS kind of Schwarzschild black hole for the 
parameter $L_0 > 1, M < 0$ as also depicted in the plot.
Now, for both the parameters $M, L_0$ being negative, the
bulk metric would be globally non-static with a naked singularity. 
On the other hand when $ L_0 <0, M> 0$, the 
second metric Eq.\ref{second} has a cosmological horizon.

Now, for $ c^2 < 1$ and $ L_0 <0, M> 0$, the metric
can  have either a RN or Schwarzschild black hole
depending upon the value of ${\cal M}_2$. Whereas for $1 >c^2 > n-1$
we can have again a RN black hole in the bulk.
But for both the cases, the bulk metric can
have naked singularity depending upon the region
of the parameter space.

\begin{figure}
\includegraphics[width=5.50in,height=2.2in]{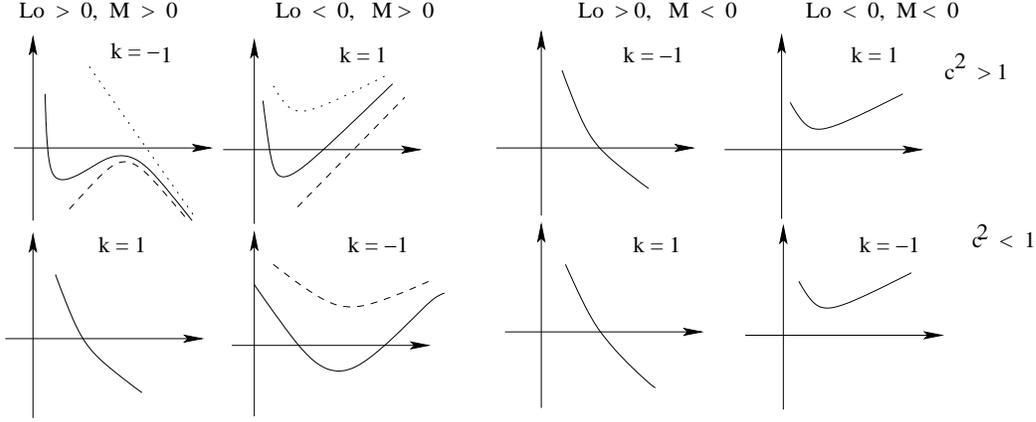}
\caption{N(r) for type-III solution
} \label{three}
\end{figure}

{\bf Type-III solution}: For $\alpha = \frac 2 {\beta(n-2)}  =  \gamma ~;~ k \ne 0$, the
metric has no solution with flat spatial section. The solution looks like
\bea
N(r)&=&  (1 + c^2)^2 r^{\frac 2 {1 + c^2}} \left[ \frac {-2 L_0}
{(n - 3)c^2 + 1}
- 2 M r^{- \frac {(n - 3) c^2 + 1}{1 + c^2}} \right] + \frac {8 \lambda^2 \Omega r^{\frac {2c^2}{1 + c^2}}}
{c^2 ((n-1)c^2 -1)} \\ 
 &-& \frac {8 \lambda \Omega r^{-\frac {(n -4)c^2}{1 +c^2}}} {c^2 \xi^{n-2} ((n-1)c^2 -1)}
\left(\sqrt{4 Q^2 + \lambda^2 R^{2n -4} } -
\frac{4 (n -2) c^2 Q^2 r^{-\frac {(n-4)c^2}{1+c^2}} }
{\lambda \xi^{n-2}((n-3)c^2 +1)} \Delta(r,Q) \right) \nonumber   \\
&&R(r) =\xi r^{\frac {c^2} {1 + c^2}} ~~~;~~~ \phi(r) = \sqrt{n-2}\left(
\phi^*_0 - \frac c {1 + c^2} log (r)\right),
\eea
where, we use the notation
\bea
\xi = \sqrt{\frac { k (n-3)}{2 L_0(1 - c^2)}}~~;~~
\Delta(r,Q) = {_2F_1}\left[\frac {(n-3)c^2 +1}{c^2(2n-4)},\frac 1 2,\frac{(3n-7)c^2 + 1}{(2n-4)c^2},
-\frac{4 Q^2 R^{(4-2n)} 
}{\lambda^2} \right].
\eea

For large $r$ limit, expression for the above solution terns out to be
\bea
\frac {N(r)|_{r\rightarrow \infty}}{(1 + c^2)^2} =\frac { -2 L_0 r^{\frac 2 {1 + c^2}}} {((n-3)c^2 +1)} 
- 2 M r^{- \frac {((n - 3)c^2 -1)}{1 + c^2}}  +
\frac {16 \Omega' Q^2 r^{-\frac {(2n-6)c^2}{1 +c^2}}}{c^2 \xi^{2n-4}((n-3)c^2+1)}+ {\cal O}(r^{\frac{(10 -4n)c^2}{1+c^2}}),
\eea
whereas in small $r$ limit, it will be
\bea
\frac {N(r)|_{r\rightarrow 0}}{(1+c^2)^2} = &-&2 {\cal M}_3 
r^{-\frac {(n-3)c^2 -1}{1 +c^2}} 
- \frac {16 \lambda \Omega' {\cal Q}_3} {c^2 \xi^{n-2}
 ((n-1)c^2-1)} r^{-\frac{(n-4)c^2}{1 +c^2}} -
\frac {2  V}{(n-1)c^2 +1} r^{\frac 2{1 +c^2}}  \nonumber \\ 
&+& \frac {8 \beta^2 \Omega' r^{\frac {2c^2}{1 +c^2}} }{c^2 ((n-1)c^2 -1)} -
\frac {8 \lambda \Omega' {\cal H}_3 r^{\frac {nc^2} {1 +c^2}}} 
{c^2 ((n-1)c^2-1)} + {\cal O}(r^{\frac{(3n-4)c^2}{1+c^2}}).
\eea
Various notations are in order
\bea 
{\cal M}_3 & =& M - \frac {16 Q^2 \Omega' (n-2)\Gamma[\frac {(3n-7)c^2 +1 }{(2n -4)c^2}]\Gamma[\frac {c^2-1}{(2n-4)c^2}]}{\sqrt{\pi} \xi^{2n -4}((n-1)c^2 -1)((n-3)c^2 +1)} \left(\frac {4 Q^2}{\lambda^2} \right)^{-\frac {((n-3)c^2 +1}{(2n -4)c^2}}\\
{\cal Q}_3 &=& Q - \frac {c^2 Q (n-2) \Gamma[\frac {(3n-7)c^2 +1} {(2n -4)c^2}]\Gamma[\frac {1-c^2}{(2n-4)c^2}]}
{((n-3)c^2 +1) 
\Gamma[\frac {(n-3)c^2 +1}{(2n -4)c^2}] \Gamma[\frac {(2n -5)c^2 + 1}{(2n -4)c^2}]} 
 \\
{\cal H}_3 &=& \frac {\lambda^2}{4 Q c^2} + \frac {(c^2-1)}{((2n-3)c^2-1)}\left(\frac {4 Q^2}{\xi^{2n-4}\lambda^2} 
\right)^{-1} (Q - {\cal Q}_3). \\
\eea

Again all the solutions are singular at $r = 0$.
The Fig.\ref{three} says the detailed asymptotic
structure of the spacetime. 
The general form of the asymptotic metric would be 
\bea \label{third}
ds^2 = r^{\frac {2c^2} {1 +c^2}}\left( - L_0 r^{\frac {2-2 c^2}{1 +c^2}}
dt^2 + d\Omega_k^2\right) + \frac 1 {L_0 r^{\frac 2 {1 +c^2}}} dr^2 .
\eea
So, it is to be noted that $L_0$ characterise the nature
of asymptotic metric for full range of $c^2$. 

The structure of the metric near $r =0$ is determined by the value
of $c^2$. When $c^2 >1$ it is the parameter ${\cal M}_3$ which
plays the role. Where as if $c^2 < 1$ then singularity
structure is determined by ${\cal Q}_3$. In this case also,
the behaviour of ${\cal M}_3$ and ${\cal Q}_3$ is same
as that of Type-II solution.

It is clear from the plots, when $L_0 <0, M>0$, we can have either
Schwarzschild or RN type black hole in bulk spacetime depending
upon the value of various parameters.
Otherwise, it has naked singularity.
\begin{figure}
\includegraphics[width=5.5in,height=1.5in]{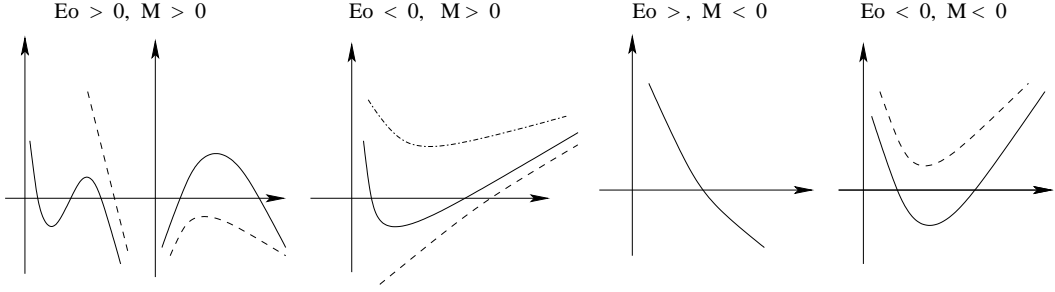}
\caption{F(R) for type-I solution.} \label{fourth}
\end{figure}

\section{Domain wall dynamics} \label{sec3}
Dynamics of the domain wall is governed by the Hubble kind of expansion equation with
respect to the domain wall observe. So, in general the equation looks like
\bea
{\dot{R}}^2 + F(R) = 0
\eea
where, $'.'$ is derivative with respect to the domain wall proper time $\tau$.
The expression for $F(R)$ is
\bea \label{pot}
F(R) = N(R)R'^2 - \frac {{\bar V}_0^2}{4 (n-2)^2} R^2,
\eea
where "prime" is derivative with respect to r.
So, the equation is like a particle moving in potential $F(R)$. So, in what
follows we will be studying different types of potential encountered by the
domain wall during the course of its motion.

{\bf Type-I potential}:
The expression for the potential is given below
\bea
F(R) = N(R) - \frac {{\bar V}_0^2}{4(n-2)^2} R^2.
\eea
The form of the potential as shown in Fig. 4 is like  
metric function $N(R)$ with an asymptotic modification by the 
domain wall tension.

Now, we will try to analyze the motion of the domain wall in various limits
of the scale factor or distance of the wall along the bulk radial co-ordinate.
 Rather than writing
 down the exact expression for the potential, it is better
 to see the limiting cases.  So, in the large $R$ limit, 
 the equation of motion would be
\bea
H^2 = \frac {{\dot{R}}^2}{R^2} = - \frac k {R^2} + E_0 + 2 M R^{-(n - 1)} -
{\cal O} (R^{-(2n-4)}),
\eea
where
\bea
E_0 =\frac {2 L_0}{(n-1)(n-2)} + \frac {{\bar V}_0^2}{4(n-2)^2}. 
\eea

For small $R$ limit,
\bea
H^2 = - \frac k {R^2} + 2 {\cal M} R^{-(n-1)} 
&+& \frac {16 \lambda {\cal Q}} {(n-1)(n - 2)} R^{-(n-2)}\nonumber \\
&+& \frac {2 {\cal V}}{(n-1)(n-2)} + \frac {{\bar V}_0^2}{4(n-2)^2}+
{\cal O}(R^{n-2}). \eea

So, from the above equation it is clear that for flat spatial section
$k =0$, with $({\cal M}_1> 0)$, the domain wall is 
radiation dominated in the early 
epoch followed by the matter domination 
and cosmological constant domination
respectively. Where as the bad feature is that during the course of its
motion, domain
wall hits the bulk singularity
at $R = 0$ for finite period of time. On the other hand, 
with the parameters being 
${\cal M}_1 <0$ but $M >0$, the domain wall starts from a finite
value of the scale factor with matter energy domination followed
by late time cosmological constant domination for $E_0 > 0$. 
Domain wall never reaches the bulk singularity in this case. 
A kind of bouncing universe scenario appears
in th domain wall world volume (recent study \cite{supratik1}). 
For the first case, the bulk space time may either be non-static 
space time with naked singularity or a Schwarzschild black hole.
For the second case, the bulk can have a de-Sitter 
horizon or it can be RN black hole depending upon the sign of 
$V_0$.
Furthermore, a special range of value of parameters exist so that the 
domain wall depicts a periodic universe between the 
two horizons of the bulk black hole background. During this stage
of evaluation, the domain wall has a finite period of inflation
near the minimum of the potential followed by the standard deceleration.
This case happens for both the cases $M >0, E_0 >0$ and $M <0, E_0 <0$
as is seen from the plot as well.

For every case, asymptotic dynamics of the domain wall
is fixed by the interplay between the bulk cosmological constant 
and the brane tension. So, if $E_0 > 0$, then domain wall undergoes
exponential inflation 
\bea
R(\tau) \propto e^{\sqrt{E_0} \tau}.
\eea

Another interesting case is $M <0, E_0 >0$, which again leads to
bouncing cosmology. So, domain wall starts collapsing from infinity
to a finite value of $R = R_0$ say. At this value, domain wall
gets repelled back to infinity by the timelike singularity in the bulk.
The value of $R_0$ may be inside the de-Sitter horizon or
the inner horizon of a RN type black hole of the background spacetime.

{\bf Type-II potential}: The domain wall strictly restricted to be of 
spatially flat metric $(k = 0)$. 
However, we note below the Hubble equation of the domain wall 
for in large $R$ limit,
\bea
H^2 = P_0 R^{-2 c^2}+ {\cal O}(R^{-(n-1 +c^2)}),
\eea
where
\bea
P_0 = \left[ \frac {2 L_0}{n-1-c^2} + \frac {{\bar V}_0^2 \Omega'}
{4(n-2)}\right] 
\eea
\begin{figure}
\includegraphics[width=5.50in,height=3.40in]{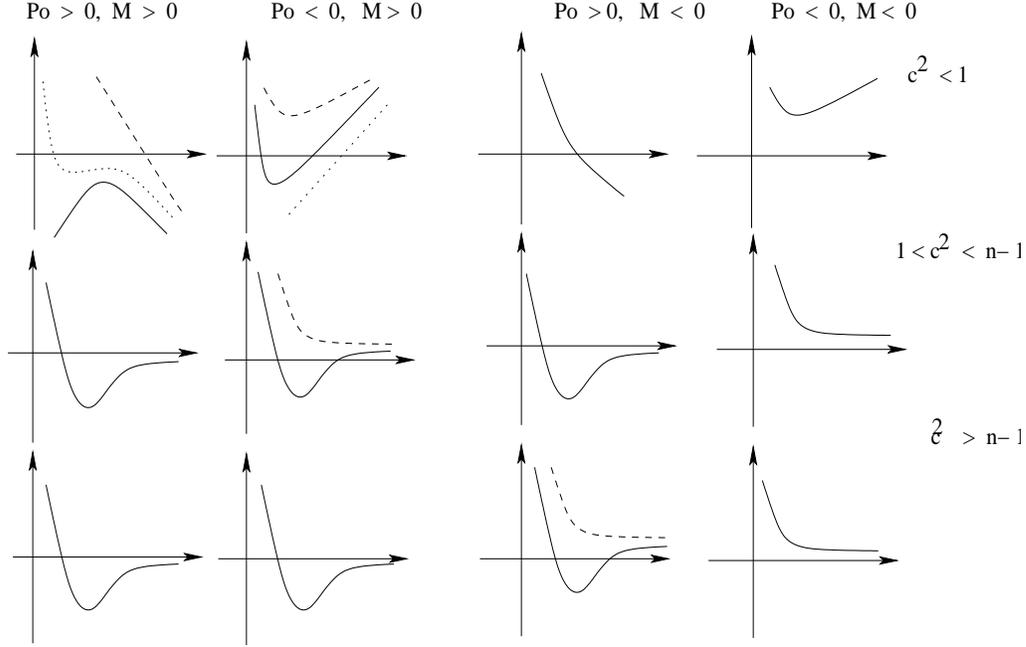}
\caption{F(R) for type-II solution. For the range $ n-1 < c^2 < n-3$,
the qualitative form of the potential is same as for $ 1 < c^2 < n-3$}. 
\label{fifth}
\end{figure}

On the other hand for small $R$ limit, the expression for the above Hubble
like equation turns out to be
\bea
H^2 = 2 {\cal M}_2 R^{-(n-1 +c^2)} &+& \frac {16 \lambda \Omega' {\cal Q}_2}
{n-1-c^2} R^{-(n-2 + 2c^2)} + \left( \frac {2 {\cal V}}{n-1-c^2} 
+ \frac {{\bar V}_0^2 \Omega'} {4(n-2)}\right) R^{-2 c^2} \nonumber\\
&+& \frac {8 \lambda \Omega' {\cal H}_2} {n-1-c^2} R^{-(n-2 - 2c^2)}
+ {\cal O}(R^{3n -6-2 c^2}).
\eea

In this case, we have different possibilities for the form of the
potential function depending upon the value of the set of 
parameters as is also seen from Fig. 5

Case-i) F(R) is positive every where. So, in this
case there is no dynamics of the domain wall.

Case-ii) F(R) is negative for finite rang of R. For $P_0 <0,M > 0$ with
$c^2 < n-1$ and $ P_0 >0, M> 0$ with $c^2 > n-1$ we have thi 
this kind of behaviour. However, in this
case, the potential has minimum which corresponds to a short
period of inflation followed by the decelerated expansion.
The domain wall has a period oscillation between two 
extreme value of the scale factor.

Case-iii) F(R) is positive for small $R$ and then negative for large $R$.
Now depending upon the value of $c$, the asymptotic structure of the potential
again is of two different types. As we note asymptotically 
$F(R) \sim -P_0 R^{2 -2 c^2}$. So, it is clear that for $P_0 > 0$,
when $c^2 > 1$, $F(R)$ grows negatively clear from the plot also.
For  $c^2 < 0$, $F(R)$ tends to zero value from negative side.
So, generally, domain wall starts from infinity and
gets repelled by the time like singularity in the bulk 
from a finite value of $R$. Asymptotically, the dynamics is
\bea
R(\tau) \propto \tau^{\frac 1 {c^2}}.
\eea
So, $c^2 <1$ gives a late time accelerated expansion of the domain wall.
For other case, expansion is decelerated. For all the cases we
get bouncing universe scenarios.

Case-iv) $F(R)$ is negative every where. In this case, the
domain wall has big-bang singularity at $R =0$. For $P_0 >0,M >0$ and
$c^2 > 1$, we have this form of the potential. Domain wall
starts collapses from infinity all the way to bulk singularity.

Case-v) $F(R)$ is negative for small value of $R$ but positive 
for large value. This occurs only for $P_0 < 0,M >0$ irrespective
of the value of $c^2$. In this potential domain wall emerges from
a the white hole region of the bulk spacetime and gets halted at finite
value of its scale factor and re-collapses into the black hole.

\begin{figure}
\includegraphics[width=5.50in,height=2.2in]{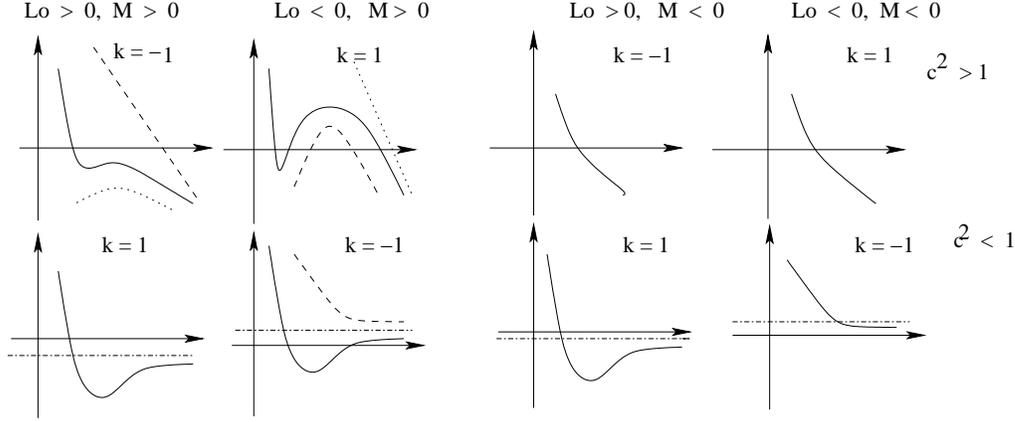}
\caption{F(R) for type-III solution.} \label{sixth}
\end{figure}

{\bf Type-III potential}: The domain wall in this 
background strictly restricted to be of
spatially non-flat metric $(k\ne 0)$. 
For large $R$ limit Hubble equation for the potential Eq.\ref{pot} 
with this solution looks like
\bea
H^2 =\frac {\xi^2 c^4}{ R^2} \left[ \frac {2 L_0}{(n-3)c^2 +1} + 
2 M \left(\frac R {\xi}\right)^{-\frac {((n-3)c^2 +1)} {c^2}}
+ \frac {{\bar V}_0^2 \Omega' \left(\frac R {\xi}\right)^{2 - \frac 2 {c^2}}}
{4(n-2)}\right] + {\cal O}(R^{-2n-6 }).
\eea
On the other hand for small $R$ limit, the expression for the above
Hubble like equation turns out to be
\bea
H^2 &=& \frac {\xi^2 c^4}{ R^2} \left[ 2 {\cal M}_3 
\left(\frac R {\xi}\right)^{-\frac {(n-3)c^2 +1}
{c^2}} + \frac {16 \lambda \Omega' {\cal Q}_3} {c^2 \xi^{n-2}
 ((n-1)c^2-1)} \left(\frac R {\xi}\right)^{-\frac{(n-4)c^2 +2}{c^2}} +
\frac {2 L_0}{(n-1)c^2 +1} \right.  \nonumber \\ 
&+& \left. \frac {{\bar V}_0^2 \Omega' 
\left(\frac R {\xi}\right)^{2 - \frac 2 {c^2}}}
{4(n-2) c^2}-\frac {8 \lambda^2 \Omega' 
\left(\frac R {\xi}\right)^{2 -\frac 2{c^2}}}{c^2 ((n-1)c^2 -1)} -
\frac {8 \lambda \Omega' {\cal H}_3 
\left(\frac R {\xi}\right)^{n-\frac 2 {c^2}}} {c^2 ((n-1)c^2-1)} \right]  + 
{\cal O}(R^{\frac{(3n-6)c^2-2}{c^2}}).
\eea

As is seen from the plot Fig.\ref{sixth}, qualitatively the
form of potential can be categorized into two
sets corresponding to the value of $c^2$. In one $c^2 >1$,
the potential tends to asymptotic negative infinity. This
leads to asymptotically power law type inflation. In the
small $R$ limit, domain wall can have bounce due to the presence
of time like bulk singularity for full $(L_0,M)$ space 
or it can go all the way to bulk attractive spacelike singularity.

In the other one $c^2 <1$, the potential is asymptotically
goes to some constant value depending on the value of $L_0$.
The constant $L_0$ is shown in the plots by dot-dashed horizontal
lines. For this, we have three kind of potential structures.
In many respects, these are of the same form 
compare to type-II potential. But main difference is the 
value of the parameter $c^2$. It acts as a inverse of that
of type-II case.

Asymptotically, the domain wall undergoes 
\bea
R(\tau) &\propto & \tau~~~~  \hbox{for $c^2 < 1$ and $L_0 >0$}\\
R(\tau) &\propto &\tau^{c^2}~~~ \hbox{for $c^2 > 1$ }.
\eea
So, for the first case, it is linear expansion with spherical
spatial section. Where as for the second case, domain wall
has power law inflation. For almost every case, domain wall
has a bounce for a finite value of scale factor.

\section{Conclusion}\label{con}
To summarize, in this report we have
tried  to study the Einstein-Born-Infeld-dilaton theory
in presence of dynamic domain walls. 
We have first tried to find out the possible background
solutions taking into account the domain wall back-reaction.
We have analytically found three different types of solutions. 
The analytical study of these various metrics
is very difficult. So, we have adopted the same
line as in \cite{chamblin,debu} by plotting all the 
metric functions and studied its structure in various
limits along the radial coordinate. 

As was mentioned earlier, BI electromagnetic field
has a critical value which is responsible for smoothing of 
pointlike singularity of the vector field. So, it is expected
to have same kind of affects on the gravitational background
in connection with the singularity at the origin. We have seen
in our previous work \cite{debu} for Maxwell electromagnetic case which
is basically $\lambda \rightarrow \infty$ limit of the present
analysis that for the first three types of solutions the singularities 
were governed by the electric charge Q
as expected. The singularities were in general timelike in those cases.
On the other hand the present analysis 
shows that for finite value of $\lambda$, singularities are somewhat
smoothed. As one can see explicitly for the first three solutions being
one to one correspondence with $\lambda \rightarrow \infty$ case, 
singularities at $r = 0$ are governed either by parameter ${\cal M}$ which
is related to mass parameter $M$ or charge Q. Importantly, depending
upon the various parameters, in these cases singularity can either be
timelike or spacelike as also clear from various plots. 
However, in order to be more explicit,
we demonstrated in the Table I that the nature of singularities appear in the
different background metric solutions for different limits of $\lambda$.
\bea
\begin{array}{c}
\hbox{Table 1: {\bf Comparison of various singularities up to sign}} \\ 
\begin{array}{|c|c|c|}
\hline
\hbox{Type of Sol}^n & \begin{array}{c} \hbox{Maxwell EM field}
\\ \lambda \rightarrow \infty \end{array} & \begin{array}{c}
\hbox{BI EM field} \\ \lambda~~\hbox{is finite}\end{array} \\
\hline
\hbox{Type-I} & Q^2 r^{-(2n-6)} & {\cal M}_1 r^{-(n-3)} \\
\hline
\hbox{Type-II} & Q^2 r^{-\frac {2n-6}{1 +c^2}} & \begin{array}{cc}
{\cal M}_2 r^{-\frac {n-3-c^2} {1 + c^2}} & \hbox{for}~c^2 < 1\\
Q r^{-\frac {n-4}{1+c^2}} &\hbox{for}~c^2 >1 \end{array} \\
\hline
\hbox{Type-III} & Q^2 r^{- \frac {(2n -6)c^2}{1+c^2}} &  \begin{array}{cc}
{\cal M}_3 r^{-\frac {(n-3)c^2-1} {1 + c^2}} & \hbox{for}~c^2 > 1\\
Q r^{-\frac {(n-4)c^2}{1+c^2}} &\hbox{for}~c^2  < 1 \end{array} \\
\hline
\end{array}
\end{array}
\eea

It may also be noted that for any value of $\lambda$
the asymptotic structure i.e. in $r \rightarrow \infty$ 
limit, the background spacetime does not depend on it at all.

Finally, after getting details of the background spacetime
we have tried to study the dynamics of the domain walls in
those bulk spacetime configurations.
In this case also, there exists specific relations
among the various coupling parameters so that one can have 
static bulk spacetime background in consistent with
the dynamic domain wall. 
As a direct consequence of drastic changes of the singularity
behaviour in our present analysis compared to $\lambda \rightarrow \infty$
case \cite{debu}, the dynamics of the domain wall near the singularity
has also modified significantly follows from various plots. 

In many cases again, we also have found to exist
inflation for finite period of proper time with
respect to domain wall world volume followed by standard decelerated
expansion phase. Another important feature is the presence
of negative energy density in domain wall scenario. 
In fact, the very presence of this negative energy density, plays the role 
of bounce for the domain wall avoiding bulk singularity.
This has recently been discussed \cite{supratik1}
for the first solution. But we have several solutions with
different asymptotic behaviour as well as near singularity structure
for the same kind of background field configurations. So,
in that respect it might be interesting to study this bounce in details for
the other non-trivial background solutions.

In the context of dark energy and dark matter in our universe,
these various kind of induced unseen energy density 
on the wall may be interesting points to study. For
example, this non-standard behavior 
may help us to construct dark matter
and dark energy \cite{dark} model building \cite{supratik}
in solving discrepancies with standard general relativity predictions
for the galaxy rotation curves \cite{galaxy}, late time acceleration of the universe \cite{acce},
gravitational lensing \cite{lensing}.
As an another possible interesting extension of this work
would be to analyze stability under
perturbation in the domain wall world volume.
An interesting point to
analyze would be whether all these types of solutions are
compatible in addition to external matter sources such as
radiation and baryonic matter, restricted to the domain
world volume.

\vspace{.1cm}
\noindent
{\bf Acknowledgment}\\
  The referee's valuable comments and suggestions are
  gratefully acknowledged.


\begin{thebibliography}{99}
\markright{Bibliography}
\bibitem{joseph} D. W. Joseph, Phys. Rev. {\bf 126}, 319 (1962).

\bibitem{akama}   K. Akama, Lect. Notes Phys.{\bf 176},
267 (1982)[hep-th/0001113]; K. Akama, Prog. Theor. Phys. {\bf 60}, 1900 (1978);
K. Akama, Prog. Theor. Phys. {\bf 78}, 184 (1987); {\bf 79}, 1299
(1988); {\bf 80}, 935 (1988); K. Akama and T. Hattori, Mod. Phys. Lett.
{\bf A15}, 2017 (2000).

\bibitem{rubakov} V.A. Rubakov and M.E. Shaposhnikov, Phys. Lett. {\bf B125}, 136 (1983);
 M. Visser, Phys. Lett. {\bf 159B}, 22 (1985).


\bibitem{other} P. Laguna-Castillo and R. A. Matzner, Nucl. Phys. {\bf B282}, 542 (1987);
E. J. Squires, Phys. Lett. {\bf B167}, 286 (1986);
G. W. Gibbons and D. L. Wiltshire, Nucl. Phys. {\bf B287}, 717 (1987);
J.M. Overduin and P.S. Wesson, Phys. Rept. {\bf 283}, 303 (1997).

\bibitem{randall} L. Randall, R. Sundrum, Phys. Rev. Lett. {\bf 83}, 3370 (1999)[hep-ph/9905221];
Phys. Rev. Lett. {\bf 83}, 4690 (1999)[hep-th/9906064].


\bibitem{dvali} G. Dvali, M. Shifman, Phys. Lett. {\bf B396}, 64 (1997)[hep-th/9612128]; Nucl.
Phys. {\bf B504}, 127 (1996)[hep-th/9611213].
\bibitem{cvetic} M. Cvetic and H. H. Soleng, Phys. Rept. {\bf 282},
159 (1997) [hep-th/9604090].
\bibitem{pol} J. Hughes, J. Liu and J. Polchinski, 
Phys. Lett. {\bf B180}, 370 (1986).

\bibitem{lukus} A. Burt A. Ovrut and D. Waldram, Phys. Rev. {\bf D61}, 023506 (2000)[hep-th/9902071],
{\it ibid} Phys. Rev. {\bf D60}, 086001 (1999)[hep-th/9806022].



\bibitem{polchinski} J. Polchinski, Phys. Rev. Lett. {\bf 75}, 4724 (1995)[hep-th/9510017].


\bibitem{kraus} P. Kraus, JHEP {\bf 9912}, 011 (1999)[hep-th/9910149].
\bibitem{csaki}  T. Nihei, Phys. Lett. {\bf B465},81 (1999), [hep-ph/9905487];
  C. Csaki, M. Graesser, C. Kolda and J. Terning, Phys. Lett. {\bf B426},34 (1999), [hep-ph/9906513];
 P. Binetruy, C. Deffayet, U. Ellwanger and D. Langlois, Phys. Lett. {\bf B477},285 (2000), 
[hep-th/9910219];
 D. Ida, JHEP {\bf 0009},014 (2000), [gr-qc/9912002];
 C. Bercelo and M. Visser, Phys. Lett. {\bf B482},183 (2000), [hep-th/0004056];
 L. Anchordoqui, C. Nunez and K. Olsen, JHEP {\bf 0010},050 (2000), [hep-th/0007064];
 P. Bowcock, C. Charmousis and R. Gregory, Class. Quant. Grav. {\bf 17},4745 (2000), [hep-th/0007177];
 C. Csaki, J. Erlich and C. Grojean, Nucl. Phys. {\bf B604},312 (2001), [hep-th/0012143];
 Y. S. Myung, hep-th/0103241.
 D. H. Coule, Class. Quant. Grav. {\bf 1}8 (2001) 4265;
 J. P. Gregory and A. Padilla, Class. Quant. Grav. {\bf 19},4071 (2002), [hep-th/0204218].
 S. Nojiri, S. D. Odintsov and S. Ogushi, hep-th/0205187.

\bibitem{israel} W. Israel, Nuovo Cimento, {\bf B44}, 1 (1966), Erratum: {\bf B48}, 463 (1967).

\bibitem{sudipto} S. Mukherji and M. Peloso, Phys. Lett. {\bf B547}, 297 (2002)[hep-th/0205180]. A. Biswas,
S. Mukherji and S. Sekhar Pal, Int. J. Mod. Phys. {\bf A19}, 557 (2004)[hep-th/0301144]; A. Biswas and S. Mukherji,
JCAP {\bf 0602}, 002 (2006)[hep-th/0507270];S. Mukherji, S. Pal,
arXiv:0806.2507 [gr-qc].

\bibitem{novello} M. Novello and S.E.Perez Bergliaffa, arXiv:0802.1634 [astro-ph] and references there in; J. L. Hovdebo and R. C. Myers, JCAP {\bf 0311}, 012 (2003) [hep-th/0308088]. C.P. Burgess, F. Quevedo, R. Raba, G. Tasinato
and I. Zavala, JCAP, {\bf 0402}, 008 (2004) [hep-th/0310122]; M. R. Setare, F. Darabi, Int. J. Mod. Phys. {\bf D16}, 1563 (2007) [hep-th/0605081].
G. De. Risi, Phys. Rev. {\bf D7}, 044030 (2008).
\bibitem{teys} E. S. Fradkin and A.A. Tseytlin, 
Phys. Lett. {\bf B160}, 69 (1985).

\bibitem{leigh}R. G. Leigh, Mod. Phys. Lett. {\bf A4}, 2767 (1989).
\bibitem{tdey} T. Kr. Dey, Phys. Lett. {\bf B595}, 
484 (2004)[hep-th/0406169]; R. Cai,
D. Pang and A. Wang, Phys. Rev. {\bf D70}, 124034 (2004)[hep-th/0410158].

\bibitem{Born-infeld} D. L. Wiltshire, 
Phys. Rev. {\bf D38}, 2445 (1988); 
D. A. Rasheed, arXiv:hep-th/9702087; M. Cataldo and A. Garcia, 
Phys. Lett. {\bf B456}, 28 (1999)[hep-th/9903257]; S. Fernando and 
D. Krug, Gen. Rel. Grav. {\bf 35}, 129 (2003)[hep-th/0306120];
T. Tamaki, JCAP {\bf 0405}, 004 (2004)[gr-qc/0310099]; S. Fernando,
Gen. Rel. Grav. {\bf 37}, 585 (2005)[hep-th/0407062]; 
S. Fernando, C. Holbrook, Int. J. Theor. Phys. {\bf 45}, 1630 (2006)
[hep-th/0501138]; E. F. Eiroa, Phys. Rev. {\bf D73} 043002 (2006)
[gr-qc/0511065]; S. Fernando, Phys. Rev. {\bf D74}, 104032 (2006)
[hep-th/0608040]; A. Sheykhi, N. Riazi, Phys. Rev. {\bf D75}, 024021 (2007)
[hep-th/0610085]; M. H. Dehghani, S. H. Hendi, A. Sheykhi and 
H. R. Sedehi, JCAP {\bf 0702}, 020 (2007)[hep-th/0611288];
X. Gao, JHEP {\bf 0711}, 006 (2007)[arXiv:0708.1226] [hep-th];
S. Yun, arXiv:0706.2046 [hep-th]; B. Chandrasekhar, H. Yavartanoo and 
S. Yun, Phys. Lett. {\bf B660}, 392 (2008)[hep-th/0611240];
A. Sheykhi, Phys. Lett. {\bf B662}, 7 (2008)[arXiv:0710.3827] [hep-th];
A. Sheykhi, arXiv:0801.4112 [hep-th]; O. Miskovic and R. Olea
arXiv:0802.2081 [hep-th]; Y. S. Myung, Y. Kim and Y. Park,
arXiv:0804.0301 [gr-qc]; Y. S. Myung, Y. Kim and Y. Park, 
arXiv:0805.0187 [gr-qc]; S. S. Yazadjiev, Phys. Rev. {\bf D72}, 044006 (2005)
[hep-th/0504152]; M.H. Dehghani, N. Bostani and S.H. Hendi,
arXiv:0806.1429 [gr-qc].


\bibitem{debu} D. Maity, arXiv:0806.2041[hep-th].
\bibitem{dil} T. Tamaki and T. Torii, Phys. Rev. {\bf D62}, 061501 (2000)
[gr-qc/0004071]; Clement and D. Galtsov, Phys. Rev. {\bf D62},
124013 (2000)[hep-th/0007228].

\bibitem{chamblin} H.A. Chamblin and  H.S. Reall, 
 Nucl. Phys. B {\bf 562}, 133 (1999); A. Chamblin, M. J. Perry 
and H. S. Reall, JHEP {\bf 9909}, 014 (1999)[hep-th/9908047].


\bibitem{0806.2481} M.H. Dehghani {\it et al} Phys. Rev.{\bf D77}, 
104025 (2008)[arXiv:0802.2637] [hep-th]; 
Z. Guo, N. Ohtaa and T. Toriib, arXiv:0806.2481[gr-qc].

\bibitem{abbot} L. F. Abbot and S. Deser, Nucl. Phys. {\bf B195}, 76 (1982);
S. W. Hawking and G. T. Horowitz, Class. Quant. Grav. {\bf 13}, 1487 (1996).

\bibitem{supratik1} S Mukherji and S. Pal, arXiv:0806.2507 [gr-qc].


\bibitem{dark} L. Bergstrom, Rept. Prog. Phys. 63, 793 (2000); F. Combes, New Astron. Rev. {\bf 46}, 755 (2002).

\bibitem{supratik} S. Pal, S. Bharadwaj and S. Kar, Phys. Lett. {\bf B609}, 194 (2005) [gr-qc/0409023];
C. G. Boehmer and T. Harko, Class. Quant. Grav. {\bf 24}, 3191 (2007) [0705.2496 [gr-qc]].
\bibitem{galaxy} J. J. Binney and S. Tremaine, Galactic Dynamics, Princeton University Press, Princeton
    (1987); M. Persic, P. Salucci and F. Stel, Mon. Not. Roy. 
    Astron. Soc. {\bf 281}, 27 (1996); 
A. Berriello and P. Salucci, Mon. Not. Roy. Astron. Soc. {\bf 323}, 285 (2001); Y. Safue and V. Rubin,
    Ann. Rev. Astron. Astrophys. {\bf 39}, 137 (2001).

\bibitem{acce} A. G. Riess et al., Astron. J. {\bf 116}, 1009 (1998);
S. Perlmutter et al., Bull. Am. Astron. Soc. {\bf 29}, 1351 (1997);
S. Perlmutter et al., Astrophys. J. {\bf 517}, 565 (1997);
J. L. Tonry et al., Astrophys. J. {\bf 594}, 1 (2003);
S. Bridle, O. Lahav, J. P. Ostriker and P. J. Steinhardt, Science, {\bf 299}, 1532 (2003);
C. Bennet et al., Astrophys. J. Suppl. Ser. {\bf 148}, 1 (2003) [astro-ph/0302207];
G. Hinshaw et al., Astrophys. J. Suppl. Ser. {\bf 148}, 135 (2003)[astro-ph/0302217;
A. Kogut et al., Astrophys. J. Suppl. Ser. {\bf 148}, 161 (2003)[astro-ph/0302213];
D. N. Spergel et al., Astrophys. J. Suppl. Ser. {\bf 148}, 175 (2003)[astro-ph/0302209].

\bibitem{lensing} P. Schneider, J. Ehlers and E. Falco, Gravitational lenses , Springer Verlag, Berlin (1992)


\end{thebibliography}
\end{document}